\documentclass[11pt,a4paper]{article}
\usepackage[utf8x]{inputenc}
\usepackage{ucs}
\usepackage{latexsym}
\usepackage{amsmath}
\usepackage{amsfonts}
\usepackage{amssymb}
\usepackage{tensor}
\usepackage{graphicx}
\usepackage{subfigure}
\usepackage{sidecap}

\newcommand{\R}{\mathbb R}

\newcommand{\doublefig}[3]
{
\begin{figure}[Htpb]
\centering
    \subfigure[]{\includegraphics[scale=0.6]{figs/rsuppq0/#1}}
    \subfigure[]{\includegraphics[scale=0.6]{figs/individ2/#1}}
\caption{#3}
\label{#2}
\end{figure}
}

\author{H{\aa}kan Andr\'{e}asson\thanks{Support by the Institut Mittag-Leffler (Djursholm, Sweden) is
 gratefully acknowledged.} and Mikael Eklund\\
Mathematical Sciences\\ University of Gothenburg\\
        Mathematical Sciences\\Chalmers University of Technology\\
        S-41296 G\"oteborg, Sweden\\
        email: hand@chalmers.se, mikekl@student.chalmers.se\\
        \ \\
        Gerhard Rein\\
        Mathematisches Institut der Universit\"at Bayreuth\\
        D-95440 Bayreuth, Germany\\
        email: gerhard.rein@uni-bayreuth.de}
\title{A numerical investigation of the steady states\\ of the spherically
       symmetric Einstein-Vlasov-Maxwell system}
\begin{document}
\maketitle

\begin{abstract}
We construct, by numerical means, static solutions of the
spherically symmetric Einstein-Vlasov-Maxwell system and
investigate various features of the solutions. This extends a
previous investigation \cite{AR1} of the chargeless case. We study the
possible shapes of the energy density profile as a function of the
area radius when the electric charge of an individual particle
is varied as a parameter. We find profiles
which are multi-peaked, where the peaks are separated either by
vacuum or a thin atmosphere, and we find that for a sufficiently
large charge parameter there are no physically meaningful
solutions. Furthermore, we investigate if the inequality
\begin{equation}\nonumber
\sqrt{M}\leq \frac{\sqrt{R}}{3}+\sqrt{\frac{R}{9}+\frac{Q^2}{3R}},
\end{equation}
derived in \cite{An2}, is sharp within the class of solutions to
the Einstein-Vlasov-Maxwell system. Here $M$ is the ADM mass, $Q$
the charge, and $R$ the area radius of the boundary of the static
object. We find two classes of solutions with this property, while
there is only one in the chargeless case. In particular we find numerical
evidence for the existence of
arbitrarily thin shell solutions to the Einstein-Vlasov-Maxwell system.
Finally, we consider one parameter families of steady states,
and we find spirals in the mass-radius diagram
for all examples of the microscopic equation of state which we
consider.
\end{abstract}

\section{Introduction}
\setcounter{equation}{0}

In this work matter is described as a large ensemble of charged
particles which interact via the gravitational and
electromagnetic fields created by the particles themselves.
All the particles have the same rest mass, normalized to $1$,
and the same charge $q_0\geq 0$.
The distribution of the particles on phase space is given by a density
function $f$. The particle ensemble is assumed to be collisionless
which implies that $f$ satisfies the Vlasov equation. Macroscopic
quantities such as mass-energy
density, pressure, and charge current, which act as source terms in
the field equations, are obtained by integrating $f$
with respect to specific weight functions.
The resulting system is called the Einstein-Vlasov-Maxwell system
and is stated in Section~\ref{sectevms}.
In the present work we construct, by numerical means,
static solutions of the asymptotically flat,
spherically symmetric Einstein-Vlasov-Maxwell system, and we
investigate three different features of the solutions. In the
chargeless case, i.e., for the spherically symmetric Einstein-Vlasov
system, a similar study has been carried out in \cite{AR1}. There
is an essential difference between these two systems concerning
known mathematical results of existence of static solutions and
their properties. In the chargeless case it is known that a wide
variety of static solutions with finite extent and finite ADM mass
exist, cf.\ \cite{An4,RR1}, and the references
therein, whereas the problem of existence of static solutions in
the charged case has not yet been studied. The mathematical
construction of steady states in the chargeless case is based on a
certain ansatz for the density function $f.$ Here this ansatz is
modified to handle the charged situation, cf.\
Section~\ref{stst}. We believe that this
constitutes a natural starting point for showing existence of
static solutions of the Einstein-Vlasov-Maxwell system,
but we do not include such an analysis here since the purpose of
the present paper is to investigate numerically three features of
static solutions which we now describe in some detail.

In Section~4 an analysis of the behavior of the energy density as
a function of area radius is carried out for different values of
the charge parameter $q_0$. Qualitatively we find a similar structure as in the
chargeless case \cite{AR1}, e.g. there are solutions with an
arbitrary number of peaks, and these peaks are separated either by
vacuum or by a thin atmosphere. The choice of charge parameter
affects in some cases the number of peaks. If the charge parameter reaches
a certain critical value the solutions break down before the energy
density vanishes.
It is natural to compare this with the Newtonian
Vlasov-Poisson system for which there is no difference in the form
of the equations whether one models a mono-charged plasma or a
gravitating system, except for the sign in front of the force field.
If one studies a charged gravitating system this sign
is positive or negative depending
on whether $q_0 <1$ or $q_0>1$. Steady
states with finite extent only exists in the former case
where the effective force field is attractive.
This is in accordance with what we
find in the relativistic situation.

In Section~5 we
investigate if there are static solutions of the
Einstein-Vlasov-Maxwell system such that the inequality
\begin{equation}\label{pD}
\sqrt{M}\leq \frac{\sqrt{R}}{3}+\sqrt{\frac{R}{9}+\frac{Q^2}{3R}},
\end{equation}
which was derived in \cite{An2}, can be saturated in the sense that the quotient of the left
and right hand side in (\ref{pD}) is arbitrarily close to one. If there
are such solutions we say that the inequality (\ref{pD}) is sharp
within this class of solutions. Here $M$ is the ADM mass, $Q$ the
total charge, and $R$ the area radius of the boundary of the static object.
It was shown in \cite{An2} that (\ref{pD}) holds for any static solution
of the Einstein-Maxwell-matter system which satisfies the energy
condition
\begin{equation}\label{energy}
p+p_T\leq\rho,
\end{equation}
where $p\geq 0$ and $p_T\geq 0$ are the radial and tangential pressures
respectively and $\rho\geq 0$ is the energy density. This condition
is satisfied by Vlasov matter.
A suitable extension of the inequality (\ref{pD})
also holds inside the static object, cf.\ \cite{An2} and Section~5.
Moreover, it was shown in \cite{An2} that the inequality is sharp, and in
particular that equality is attained by infinitely thin shell
solutions. The method of proof in \cite{An2} is quite general and
applies to any matter model for which (\ref{energy}) holds.
However, the solution constructed in the analysis leading
to sharpness has features which solutions of the Einstein-Vlasov-Maxwell system
do not have. Hence, one motivation for the present study is
to investigate if sharpness of the inequality can be attained
by solutions when a real matter model is chosen so that the system of
equations includes a matter field equation, in the
case at hand the Vlasov equation. In
contrast to the chargeless case it is not known if there are
solutions other than infinitely thin shells which saturate
(\ref{pD}), and it is not known if arbitrarily thin shell solutions
do exist for the Einstein-Vlasov-Maxwell system.

In the uncharged case more is known. It follows from \cite{An3}
that infinitely thin shell solutions are {\em unique} in
saturating the inequality
(\ref{pD}) for $Q=0$, i.e., the inequality
\[
M\leq \frac{4R}{9}.
\]
Since an infinitely thin shell solution is not a regular solution
of the Einstein-matter system this statement should be interpreted
in the sense that a sequence of regular solutions tending to an
infinitely thin shell will in the limit give equality. Moreover,
it is known \cite{An4} that regular
arbitrarily thin shell solutions of the Einstein-Vlasov system do
exist, which then in particular implies that there are steady
states to this system such that $M/R$ is arbitrarily close to $4/9.$

In the present work we find numerical evidence for answering the
issues raised above. Indeed, we construct arbitrarily thin shell
solutions to the Einstein-Vlasov-Maxwell system, which saturate
the inequality (\ref{pD}) in the limit. Moreover, in contrast to
the uncharged case we also find another type of solutions which
saturate the inequality. These solutions have the feature that $M,Q$, and
$R$ are all equal; they represent an extremal object.
The latter property may be of interest in fundamental black hole
physics, cf.\ \cite{ARo,GR,FH}.

In Section 6 the third and final property is investigated,
namely the relation between the ADM
mass and the outer area radius of a one parameter family of steady
states to the Einstein-Vlasov-Maxwell system. The one parameter
family is obtained by prescribing the way in which $f$ depends on
the local energy and the angular momentum,
which we call the microscopic equation of state. We find numerical
support for mass-radius spirals for all examples of the microscopic
equation of state which we investigate.
This agrees with the result in \cite{AR1} in
the chargeless case, but on the other hand it differs from the
Newtonian situation where the presence of such spirals heavily
depends on the microscopic equation of state. In
\cite{HRU,Ma} the question of which equations
of state in the fluid case give rise to spirals is investigated.

To conclude this introduction we mention \cite{ARo,dFSY2,GR}
where numerical
investigations of solutions to the Einstein-Maxwell system are
carried out using other conditions on the matter model. These
studies focus on the issue of bounding $M$ in terms of $Q$ and $R.$
\section{The spherically symmetric Einstein-Maxwell-Vlasov system}
\label{sectevms}
\setcounter{equation}{0}
We choose general local coordinates $x^\alpha$ on the spacetime manifold
and we denote by $p^\alpha$ the corresponding canonical momenta;
Greek indices always run from $0$ to $3$ and Latin ones from $1$ to $3$.
We assume that $x^0 = t$
is a timelike coordinate and that $p^0$ can be expressed by $p^i$ through the condition that
all the particles have rest mass normalized to $1$:
$g_{\alpha \beta} p^\alpha p^\beta = -1$.
Then the Einstein-Vlasov-Maxwell system takes the following form:
\begin{gather}
F\indices{^{\alpha\beta}_{;\alpha}} = 4 \pi\,J^{\beta}, \label{eq:maxwellpart1}\\
F\indices{_{\beta\gamma;\alpha}} + F\indices{_{\gamma\alpha;\beta}} +
F\indices{_{\alpha\beta;\gamma}} = 0, \label{eq:maxwellpart2}\\
R_{\alpha\beta} - \dfrac{1}{2}g_{\alpha\beta}R = 8 \pi(T_{\alpha\beta} +
\tau_{\alpha\beta}),
\label{eq:einstein}\\
\partial_{t}f + \dfrac{p^i}{p^0}\partial_{x^{i}}f -
\dfrac{1}{p^0}(\Gamma_{\alpha\beta}^{i}p^{\alpha}p^{\beta}
+ q_{0}p^{\alpha}F\indices{_\alpha^i})\partial_{p^{i}}f = 0,
\label{eq:vlasov}
\end{gather}
where
\begin{gather}
T_{\alpha\beta}
= -\int_{\mathbb{R}^{3}}p_{\alpha}p_{\beta}\sqrt{|g|}\dfrac{d^{3}p}{p_{0}}, \\
\tau_{\alpha\beta}
= \frac{1}{4\pi}
\left(F\indices{_{\alpha}^{\gamma}}F_{\beta\gamma}
- \dfrac{g_{\alpha\beta}}{4}F_{\gamma\nu}F^{\gamma\nu}\right), \\
J^{\beta} = q_{0}\int_{\mathbb{R}^3}p^{\beta}f\sqrt{|g|}\dfrac{d^{3}p}{p_{0}}.
\end{gather}
Here \eqref{eq:maxwellpart1} and \eqref{eq:maxwellpart2}
are the Maxwell equations,
\eqref{eq:einstein} are the Einstein equations, \eqref{eq:vlasov} is
the Vlasov equation, and $F_{;\alpha}$ denotes the covariant derivative.

We consider this system under the assumption of spherical symmetry.
Hence the metric, expressed in Schwarzschild coordinates, takes the form
\begin{equation}
ds^2 = -e^{2\mu(t,r)}dt^2 + e^{2\lambda(t,r)}dr^2 + r^2(d\theta^2 + \sin^2\theta d\varphi^2),
\end{equation}
where
\begin{equation}
t \geq 0,\ r \geq 0,\ \theta \in [0, \pi],\ \varphi \in [0, 2\pi].
\end{equation}
For the metric to approach that of Minkowski space as $r$ goes to infinity,
the boundary conditions
\begin{equation}
\lim_{r \to \infty}\lambda(t,r) = \lim_{r \to \infty}\mu(t,r) = 0
\label{eq:boundcond}
\end{equation}
are imposed. Furthermore, the condition
\begin{equation}
\lambda(t, 0) = 0
\end{equation}
ensures a regular center. We introduce the corresponding
Cartesian coordinates
$x = (x^1,x^2,x^3) = r (\sin\theta \cos\varphi,\sin\theta \sin
\varphi,\cos\theta)$
and find that
\[
p_0 = - e^\mu \sqrt{1+|p|^2 + (e^{2\lambda} -1) \left(\frac{x\cdot p}{r}\right)^2}.
\]
Here $x\cdot p$ denotes the Euclidean scalar product of the vectors
$x=(x^1,x^2,x^3),\ p=(p^1,p^2,p^3)$, and $|\cdot|$ denotes the
Euclidean norm on $\R^3$.
For a spherically symmetric electric field of the form
\begin{equation*}
E^i = \varepsilon\dfrac{x^{i}}{r}
\end{equation*}
the non-zero components of the electromagnetic field-strength tensor are
\begin{equation*}
F^{0i} = e^{-\mu}\varepsilon\dfrac{x^{i}}{r}.
\end{equation*}
Let
\[
w = \frac{x\cdot v}{r},\qquad L = |x\times v|^2 = r^2(|v|^2 - w^2),
\]
where
\[
  v^{i} = p^{i} + (e^{\lambda} - 1)\dfrac{x\cdot p}{r}\dfrac{x^{i}}{r};
\]
the variables $w$ and $L$ can be viewed as the momentum in the radial
direction and the square of the angular momentum, respectively,
expressed in a suitable frame.
The system now reads
\begin{gather}
  \label{eq:y1_1}
  q' = 4\pi r^{2}\rho_{q}, \\
  \label{eq:y2_1}
  e^{-2\lambda}(2r\lambda' - 1) + 1 = 8\pi r^{2}\rho + \dfrac{q^{2}}{r^{2}}, \\
  \label{eq:y3_1}
  e^{-2\lambda}(2r\mu' + 1) - 1 = 8\pi r^{2}p - \dfrac{q^{2}}{r^{2}},
\end{gather}
\begin{eqnarray}\label{eq:evm_trwL}
&&
\partial_{t}f + e^{\mu - \lambda}\dfrac{w}{\sqrt{1 + w^2 + L/r^2}}\partial_{r}f
- \biggl(e^{\mu - \lambda}\mu'\sqrt{1 + w^2 + L/r^2} + \dot\lambda w \nonumber \\
&&
\qquad\qquad\qquad
- e^{\mu}\dfrac{q_{0}q}{r^2}
- e^{\mu - \lambda}\dfrac{L}{r^3 \sqrt{1 + w^2 + L/r^2}}\biggr)\partial_{w}f = 0,
\end{eqnarray}
where
\begin{gather}
\rho = \dfrac{\pi}{r^{2}}\int_{-\infty}^\infty\int_0^\infty\sqrt{1+w^2+L/r^2}\,f\, dL\,dw, \\
p = \dfrac{\pi}{r^{2}}\int_{-\infty}^\infty\int_0^\infty
\dfrac{w^{2}}{\sqrt{1+w^2+L/r^2}}\,f\, dL\,dw,\\
\rho_{q} = q_{0}e^{\lambda}
\dfrac{\pi}{r^{2}}\int_{-\infty}^\infty\int_0^\infty\,f\, dL\,dw.\label{rhoq}
\end{gather}
Here a prime or dot denotes the derivative with respect to $r$
or $t$, respectively, $\rho_q = \rho_q(t,r)$ is the charge
density, $q = q(t, r)$ is the charge contained in the ball
with area radius $r$ about the origin, $\rho = \rho(t,r)$ is the energy density as
defined when no charge is present, $p = p(t,r)$ is the radial
pressure, and  the modulus of the electric field is given by
\[
\varepsilon = e^{-\lambda} \frac{q}{r^2}.
\]
\section{Constructing static solutions to the spherically symmetric Einstein-Vlasov-Maxwell system}
\label{stst}
\setcounter{equation}{0}
In this paper we are interested in static solutions, so \eqref{eq:evm_trwL} reduces to
\begin{equation}\label{eq:evm_rwL}
w\partial_{r}f + \left(e^{\lambda}\dfrac{q_{0}q}{r^2}\sqrt{1 + w^2 + L/r^2} + \dfrac{L}{r^3}
- \mu'(1 + w^2 + L/r^2)\right)\partial_{w}f = 0,
\end{equation}
where $f = f(r,w,L)$, $\mu = \mu(r)$ and $\lambda = \lambda(r)$.
Due to spherical symmetry the quantity $L$ is conserved along characteristics
of the Vlasov equation, and so is the particle energy $E$ defined as
\begin{equation} \label{edef}
E = e^{\mu}\sqrt{1 + w^{2} + L/r^{2}} -
q_{0}\int_{0}^{r}e^{2\lambda(\eta) + \mu(\eta)}\varepsilon(\eta) d\eta.
\end{equation}
Hence any density function of the form
\begin{equation} \label{fansatz}
f(r, w, L) = \Phi(E, L)
\end{equation}
satisfies the static Vlasov equation \eqref{eq:evm_rwL}.
In order to motivate (\ref{edef}) we combine
the electric potential $\phi_E$ and the magnetic vector potential
$A$ into a four-vector
\[
\kappa^0 = \phi_E,\
\kappa^i = A^i.
\]
The electromagnetic field-strength tensor can be derived as
\[
F_{\alpha\beta} = \kappa_{\alpha;\beta} - \kappa_{\beta;\alpha}
= \partial_{\beta}\kappa_{\alpha} - \partial_{\alpha}\kappa_{\beta}.
\]
In particular,
\[
F_{0r} = -e^{2\lambda + \mu}\varepsilon = -\partial_{r}\kappa_{0},
\]
i.e., with the electric potential taken to be zero at $r = 0$,
\[
\kappa_{0} = \int_{0}^{r}e^{2\lambda(\eta) + \mu(\eta)}\varepsilon(\eta) d\eta.
\]
From this we get the particle energy
$E = -(p_{0} + q_{0}\kappa_{0})$ as defined in (\ref{edef}).

The ansatz (\ref{fansatz}) is a generalization to
the charged case of the standard ansatz for the Einstein-Vlasov system
which is obtained for $q_0=0$,
and to the best of our
knowledge it has not appeared in the literature. Equations
\eqref{eq:y1_1}--\eqref{eq:y3_1} can be rewritten as the system of
ODE's
\begin{gather}
\label{eq:y1_2}
\frac{d}{dr} q = 4 \pi r^{2}\rho_{q} \\
\label{eq:y2_2}
\frac{d}{dr} \left(r e^{-2\lambda}\right) = 1 - 8\pi r^{2}\rho - \dfrac{q^{2}}{r^{2}} \\
\label{eq:y3_2}
\frac{d}{dr} \left(r e^{2\mu}\right)
= e^{2(\mu + \lambda)}\left(1 + 8\pi r^{2}p - \dfrac{q^{2}}{r^{2}}\right),
\end{gather}
where the quantities $\rho_q, \rho$, and $p$ are now functionals of
$q, \lambda$, and $\mu$.
In order to obtain a steady state with finite ADM mass and finite
extension we prescribe some cut-off
energy $E_0>0$ and assume that $\Phi(E, L) = 0$ for $E > E_0$.
Taking this into account,
\begin{gather*}
\rho = \dfrac{2\pi}{r^{2}}\int_0^{w_\mathrm{max}}
\int_0^{L_\mathrm{max}}\sqrt{1+w^2+L/r^2}\Phi\, dL\,dw, \\
p =
\dfrac{2\pi}{r^{2}}\int_0^{w_\mathrm{max}}
\int_0^{L_\mathrm{max}}\dfrac{w^{2}}{\sqrt{1+w^2+L/r^2}}
\Phi\, dL\,dw, \\
\rho_{q} = q_{0}e^{\lambda}\dfrac{2\pi}{r^{2}}
\int_0^{w_\mathrm{max}}\int_0^{L_\mathrm{max}}\Phi\, dL\,dw,
\end{gather*}
where the upper limits
\begin{equation*}
w_\mathrm{max} = \left(e^{-2\mu}\left(E_0 + q_0\int_{0}^{r}e^{2\lambda(\eta) +
\mu(\eta)}\varepsilon(\eta) d\eta\right)^2 - 1\right)^{\frac{1}{2}} \\
\end{equation*}
and
\begin{equation*}
L_\mathrm{max} = r^2\left(e^{-2\mu}\left(E_0 +
q_0\int_{0}^{r}e^{2\lambda(\eta) +
\mu(\eta)}\varepsilon(\eta) d\eta\right)^2 - w^2 - 1\right)
\end{equation*}
follow from the condition $E < E_0$. Since $\varepsilon(0) = \lambda(0) = 0$,
\eqref{eq:y1_2}--\eqref{eq:y3_2} can be solved if
$\mu(0)$, $q_0$, and $E_0$ are specified. In order to continue we introduce
\begin{equation} \label{eq:ydef}
y_1 = e^\lambda \varepsilon = \dfrac{q}{r^2}, \
y_2 = e^{2\lambda}, \
y_3 = \frac{e^{2\mu}}{E_0^2}.
\end{equation}
By making the ansatz $\Phi(E, L) = \phi(E/E_0, L)$,
\eqref{eq:y1_2}--\eqref{eq:y3_2} turn into
\begin{gather}
\label{eq:y1}
\frac{d}{dr} \left(r^2y_1\right) = 4\pi r^{2}\rho_{q}, \\
\label{eq:y2}
\frac{d}{dr} \left(\frac{r}{y_2}\right) = 1 - r^{2}\left(8\pi\rho + y_1^2\right), \\
\label{eq:y3}
\frac{d}{dr} \left(r y_3\right)
= y_2 y_3\left(1 + r^{2}\left(8\pi p - y_1^2\right)\right),
\end{gather}
where
\begin{gather}
\label{eq:matquantrho}
\rho = \dfrac{2\pi}{r^{2}}\int_0^{w_\mathrm{max}}
\int_0^{L_\mathrm{max}}\sqrt{1+w^2+L/r^2}\phi\, dL\,dw, \\
\label{eq:matquantp}
p = \dfrac{2\pi}{r^{2}}\int_0^{w_\mathrm{max}}
\int_0^{L_\mathrm{max}}\dfrac{w^{2}}{\sqrt{1+w^2+L/r^2}}\phi\, dL\,dw, \\
\rho_{q} = q_{0}\sqrt{y_2}\dfrac{2\pi}{r^{2}}
\int_0^{w_\mathrm{max}}\int_0^{L_\mathrm{max}}\phi\, dL\,dw,
\label{eq:matquantrhoq}
\end{gather}
with upper limits for the integrals given by
\begin{gather*}
w_\mathrm{max} = \left(\frac{1}{y_3}\left(1 + q_0\int_{0}^{r}
\sqrt{y_2(\eta)y_3(\eta)}y_1(\eta) d\eta\right)^2 - 1\right)^{\frac{1}{2}}, \\
L_\mathrm{max} = r^2\left(\frac{1}{y_3}\left(1 + q_0\int_{0}^{r}
\sqrt{y_2(\eta)y_3(\eta)}y_1(\eta) d\eta\right)^2 - w^2 - 1\right).
\end{gather*}
Thus, since
\[
\frac{E}{E_0} = \sqrt{y_3}\sqrt{1 + w^{2} +
\dfrac{L}{r^{2}}} - q_{0}\int_{0}^{r}\sqrt{y_2(\eta)y_3(\eta)}y_1(\eta) d\eta
\]
the dependence on $E_0$ is eliminated from the system of ODE's
\eqref{eq:y1}--\eqref{eq:y3}, and the equations can be solved after
specifying merely $y_3(0)$ and $q_0$. Notice that from the boundary
conditions \eqref{eq:boundcond} we require that $e^{2\mu(r)} \to 1$ as $r \to
\infty$
so that the limit of $y_3$ at infinity and (\ref{eq:ydef}) determine $E_0$.

For almost all numerical solutions presented below
an ansatz of the form
\begin{equation}
\label{eq:numansatz}
\phi(\eta,L) = (1-\eta)_+^k(L-L_0)_+^l,
\end{equation}
where $k \geq 0,\ l > -1/2,\ k < 3l + 7/2,\ L_0 \geq 0$,
has been used. Here $x_+ = \max\{x,0\}$.
Our code does allow for a wide variety of
ansatz functions, but the qualitative behavior is similar in all
cases we have tried, cf.\ the end of Section~6.
When using an ansatz with a cut-off
$L_0>0$ for the square of the angular momentum as in
\eqref{eq:numansatz}, then at any point with
\begin{equation*}
r < \sqrt{L_0\left(\dfrac{\left(1+q_0\int_{0}^{r}
\sqrt{y_2(\eta)y_3(\eta)}y_1(\eta) d\eta\right)^2}{y_3}-1\right)^{-1}}
\end{equation*}
the matter quantities \eqref{eq:matquantrho}--\eqref{eq:matquantrhoq} are zero.
In particular the matter quantities will be zero and the functions
$y_1,y_2,y_3$ will be constant for $r < R_0$ with
\[
R_0 = \sqrt{\dfrac{L_0}{y_3(0)^{-1} -1}}.
\]
The integrals in the matter quantities are calculated using the
piecewise Simpson's rule, and the differential equations are
solved, from $r = R_0$ radially outwards, using the Euler
method with a variable step length $h_n$, given at a point $r_n$
as
\[
h_n = \dfrac{h_\mathrm{max}}{\ln{(e + |\rho''_n|)}},
\]
where $\rho''_n$ denotes the second derivative with respect
to $r$ at $r = r_n$, $h_\mathrm{max}$ is an appropriately chosen maximum
step length, and $e$ is Euler's number. This implies that the code
resolves those regions more finely where $\rho$ varies rapidly.
In the chargeless case the integrals in
\eqref{eq:matquantrho} and \eqref{eq:matquantp}
can be carried out explicitly for suitable ansatz functions $\phi$,
cf.~\cite{AR1}, and we used the corresponding code to test the
present one where these integrals are evaluated numerically.
\doublefig{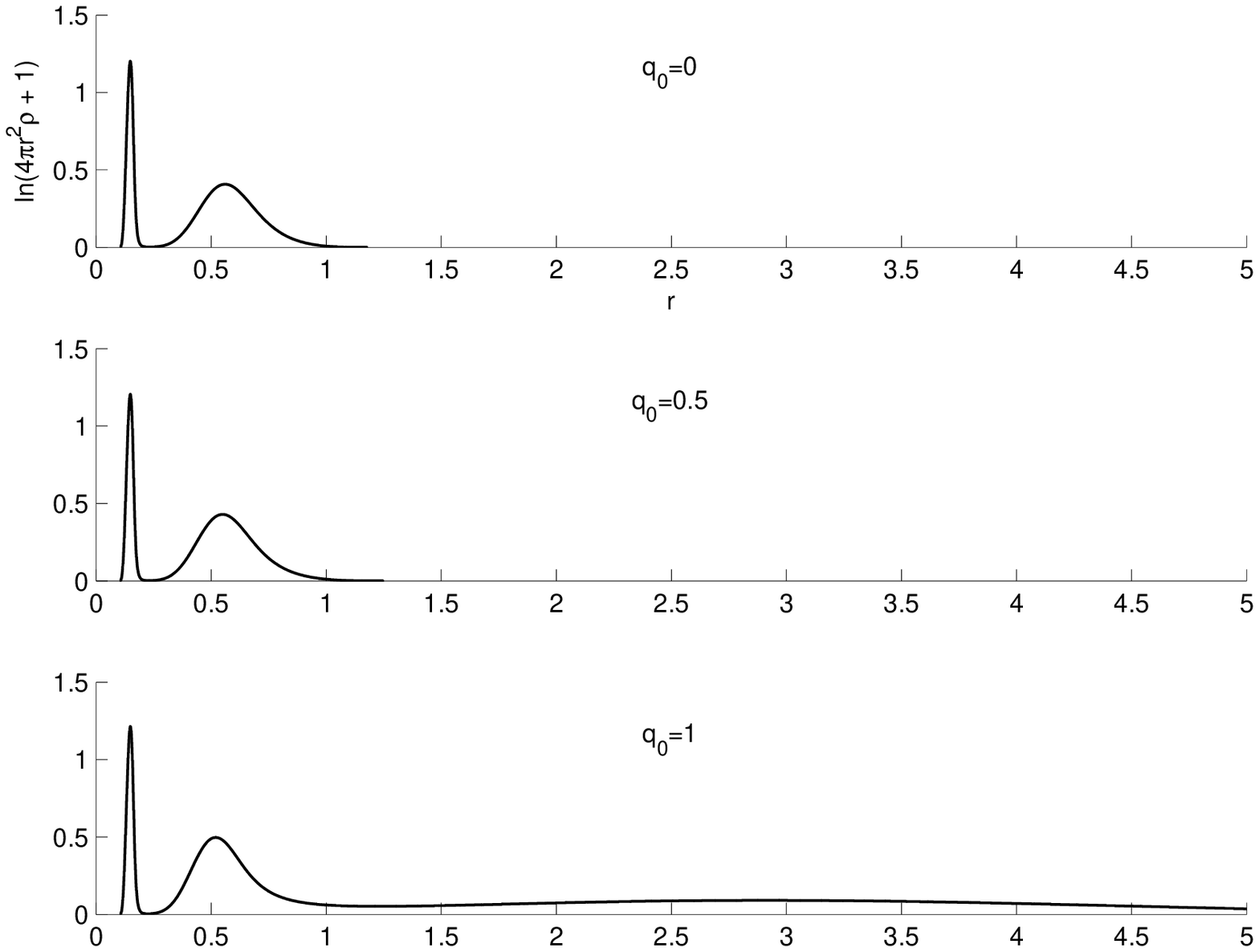}{fig:doublefig11}{Shell configurations, strictly increasing $R$}
\section{Characterization of steady states}
\label{section:charss}
\setcounter{equation}{0}
In this section we study the behavior of the uncharged energy density $\rho$
as a function of the area radius
when the parameter for electric charge is varied. We find profiles
which are multi-peaked, where the peaks are separated either by
vacuum or a thin atmosphere. In this respect our results are
qualitatively similar to the chargeless case, but
the value of the charge parameter $q_0$ does affect
the number of peaks in some cases. On the other hand,
we find that for a sufficiently
large value of the charge parameter there are no physically meaningful
solutions. This is intuitively clear in view of the Newtonian case where
no steady states of finite mass exist when the effective force is repulsive.

The steady states for charged matter largely follow the same basic
structures as noted in \cite{AR1} for the non-charged case,
i.e., we get steady states with support for $\rho$ in $[R_0, R]$,
$R_0 > 0$, called shell configurations, and states where $R_0 =
0$, called ball configurations.

\begin{figure}[Htpb]
\centering
    \includegraphics[scale=0.6]{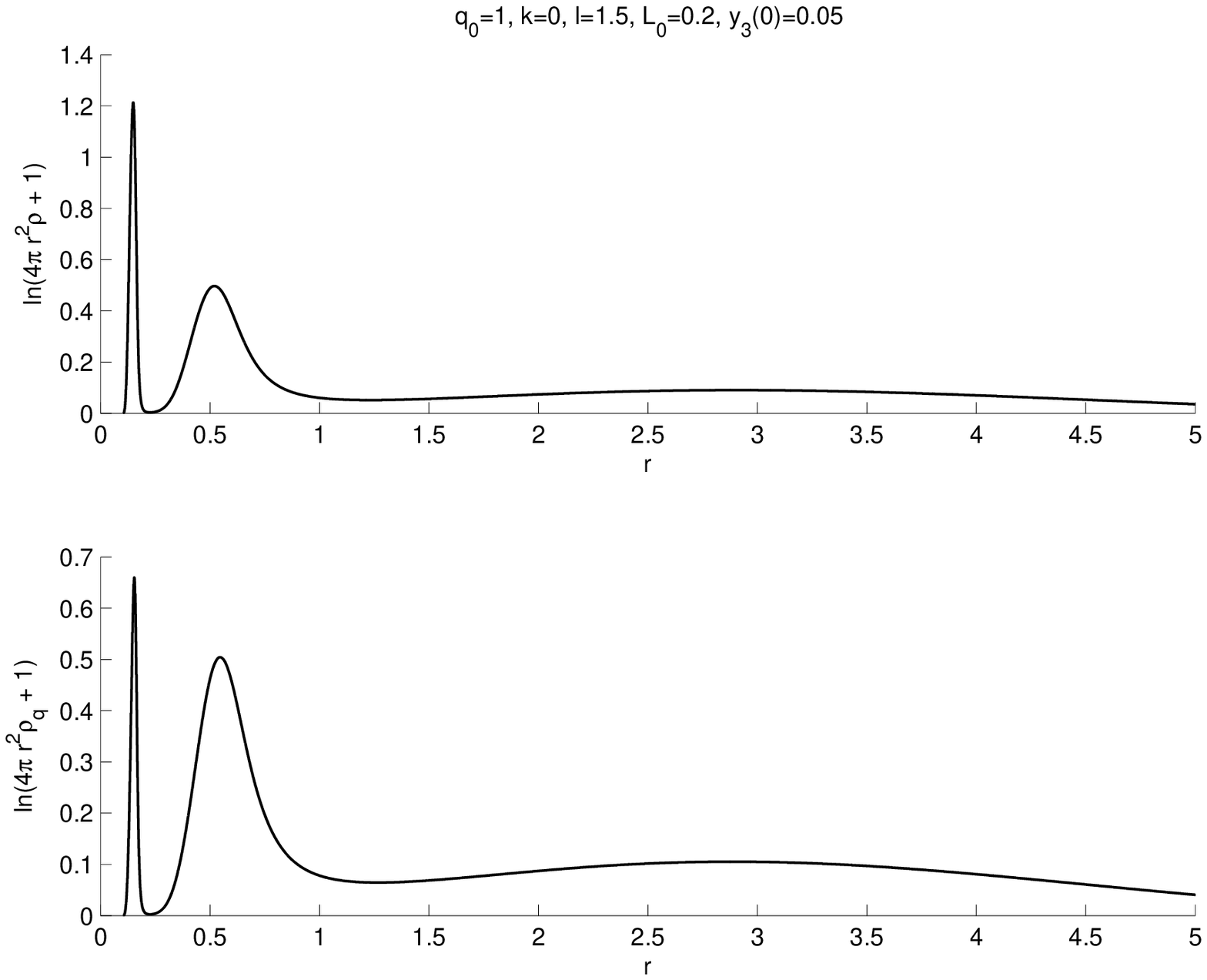}
\caption{$\ln{(4\pi r^2\rho+1)}$ and $\ln{(4\pi r^2\rho_q+1)}$ for a shell configuration}
\label{fig:rhoq}
\end{figure}

\doublefig{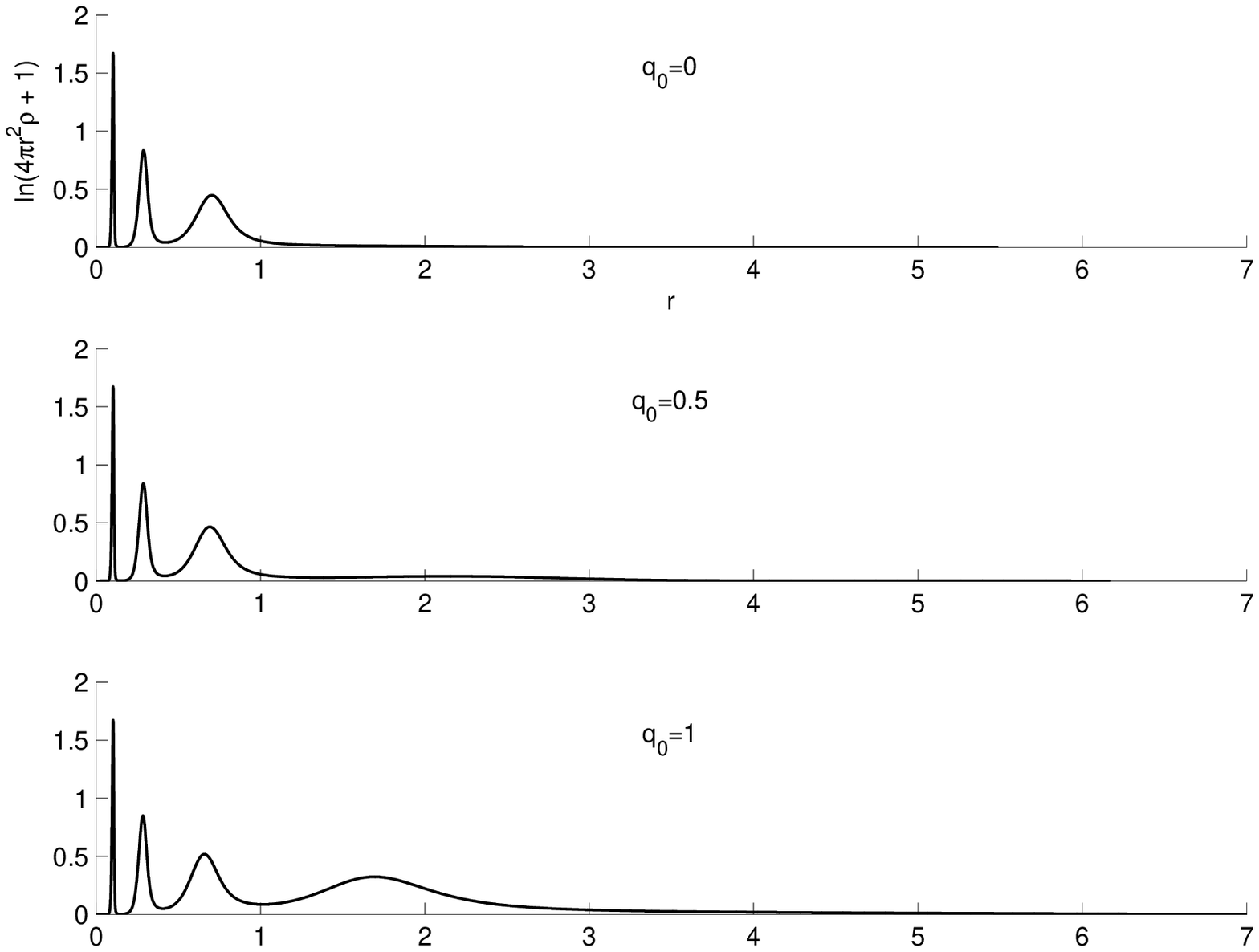}{fig:doublefig08}{Ball configurations, decreasing $R$}
\doublefig{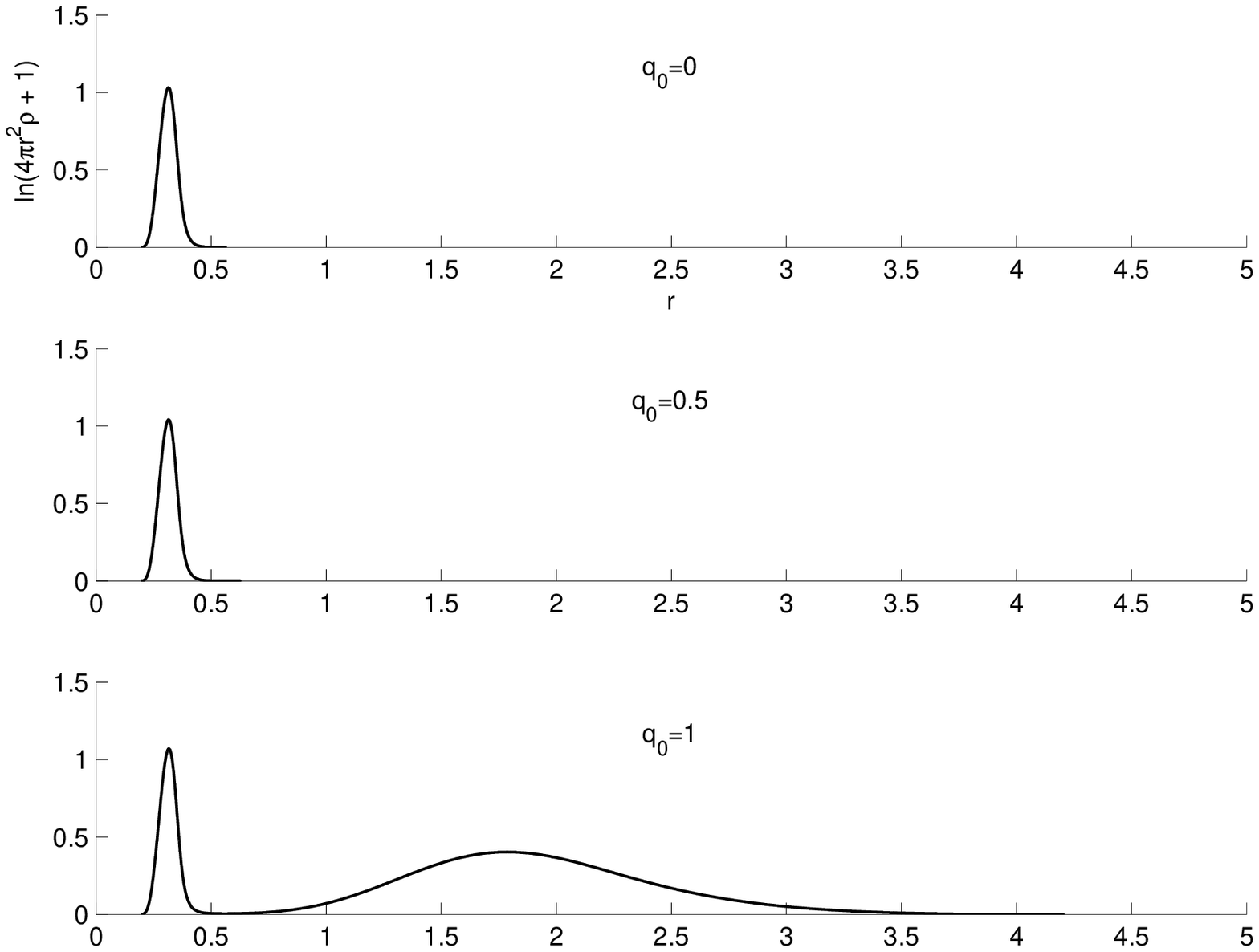}{fig:doublefig12}{Shell
  configurations, decreasing $R$}

To visualize the steady states, for each quadruple
$(k,l,L_0,y_3(0))$ a figure containing two subfigures is
presented. Subfigure (a) shows starting points, stopping points,
i.e., the inner and outer boundaries of the matter shells, as
well as maxima of $\rho$ for $q_0 \in [0,q_{0,c}[$, while
subfigure (b) shows individual solutions for three values of
$q_0$, namely $q_0 = 0, 0.5, 1.0$. The starting points in
subfigure (a) are shown as dashed lines, the stopping points as
solid lines and peaks are represented by dotted lines. The plots
complement each other, the plot for the starting and stopping
points contains little information on the shape of the steady
states, while vacuum regions can be difficult to notice in the
plots for individual solutions. To remedy the fact that peaks for
$\rho$ closer to $r = 0$ are in general much larger in magnitude
than peaks farther out, $\ln(4\pi r^2\rho + 1)$ rather than $\rho$
has been plotted against $r$ for the individual solutions. It
should be borne in mind that this has the effect that the
positions of (and in one case also number of) maxima in subfigures
(a) and (b) differ.
\doublefig{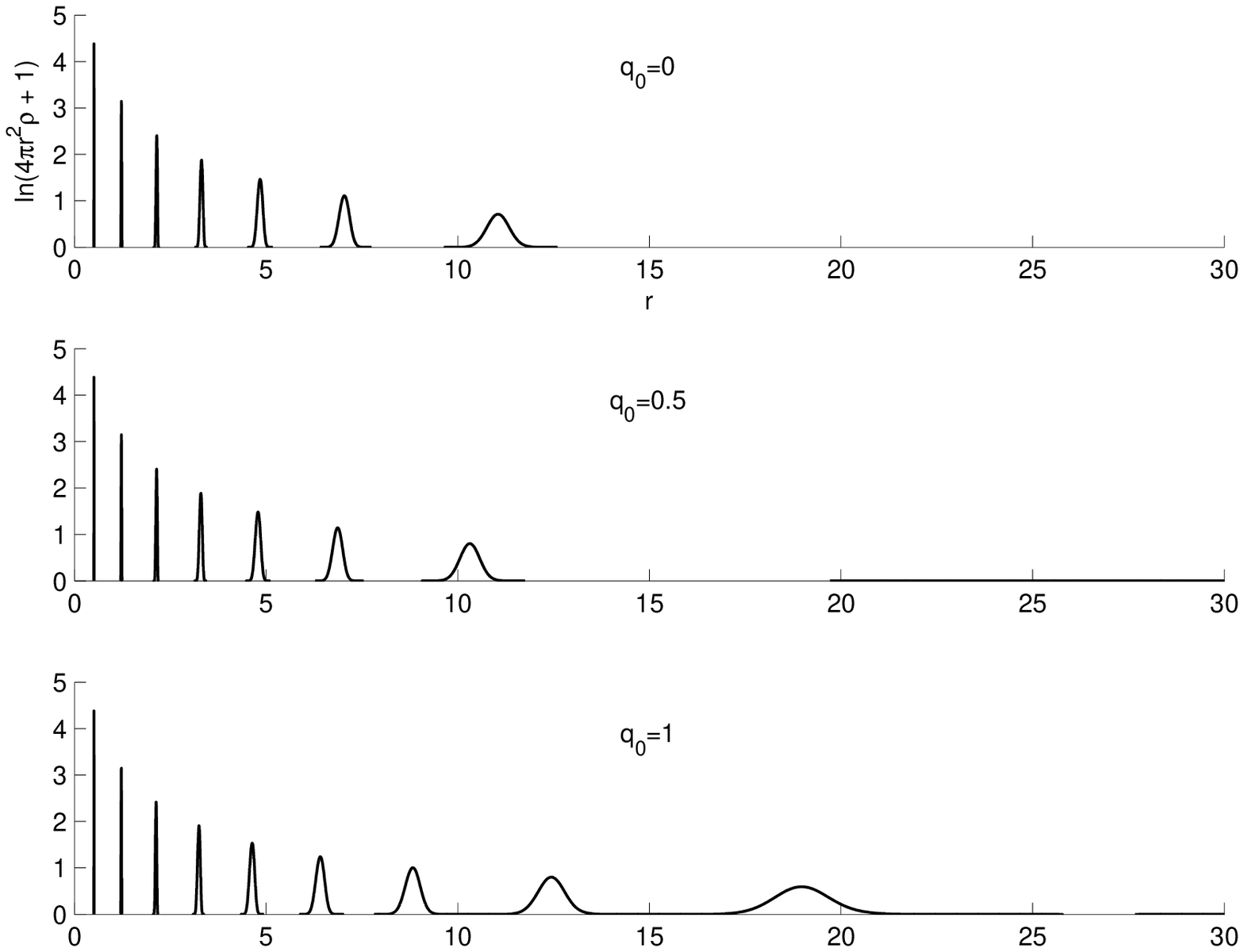}{fig:doublefig03}{Multi-shell configurations, extra shells appears}

The most noticeable effect of changing the particle charge $q_0$ is that as
$q_0$ approaches a critical value $q_{0,c}$, the outer radius $R$ of the support
for $\rho$ increases dramatically, as can be seen in
Figures~\ref{fig:doublefig11} and \ref{fig:doublefig08}--\ref{fig:doublefig03}.
Intuitively, this is expected, since for some value of $q_0$,
the repulsive forces between individual particles from electric charge
balances the
attractive forces of gravity. For values of $q_0$ larger than $q_{0,c}$, the
numerical solution breaks down since at some point $y_2(r)$ approaches
infinity. Thus we cannot have arbitrarily large particle charge and still obtain
a solution. In Figure~\ref{fig:rhoq} the behaviour of $\ln{(4\pi r^2\rho+1)}$ and
$\ln{(4\pi r^2\rho_q+1)}$ is displayed. Recall here the definition of $\rho_q$ in (\ref{rhoq}). As can
be seen the profiles of these quantities are similar. This observation applies to other cases
in this paper as well which is the reason why we only focus on $\ln{(4\pi r^2\rho+1)}.$

\begin{figure}[Htpb]
\centering
    \includegraphics[scale=0.6]{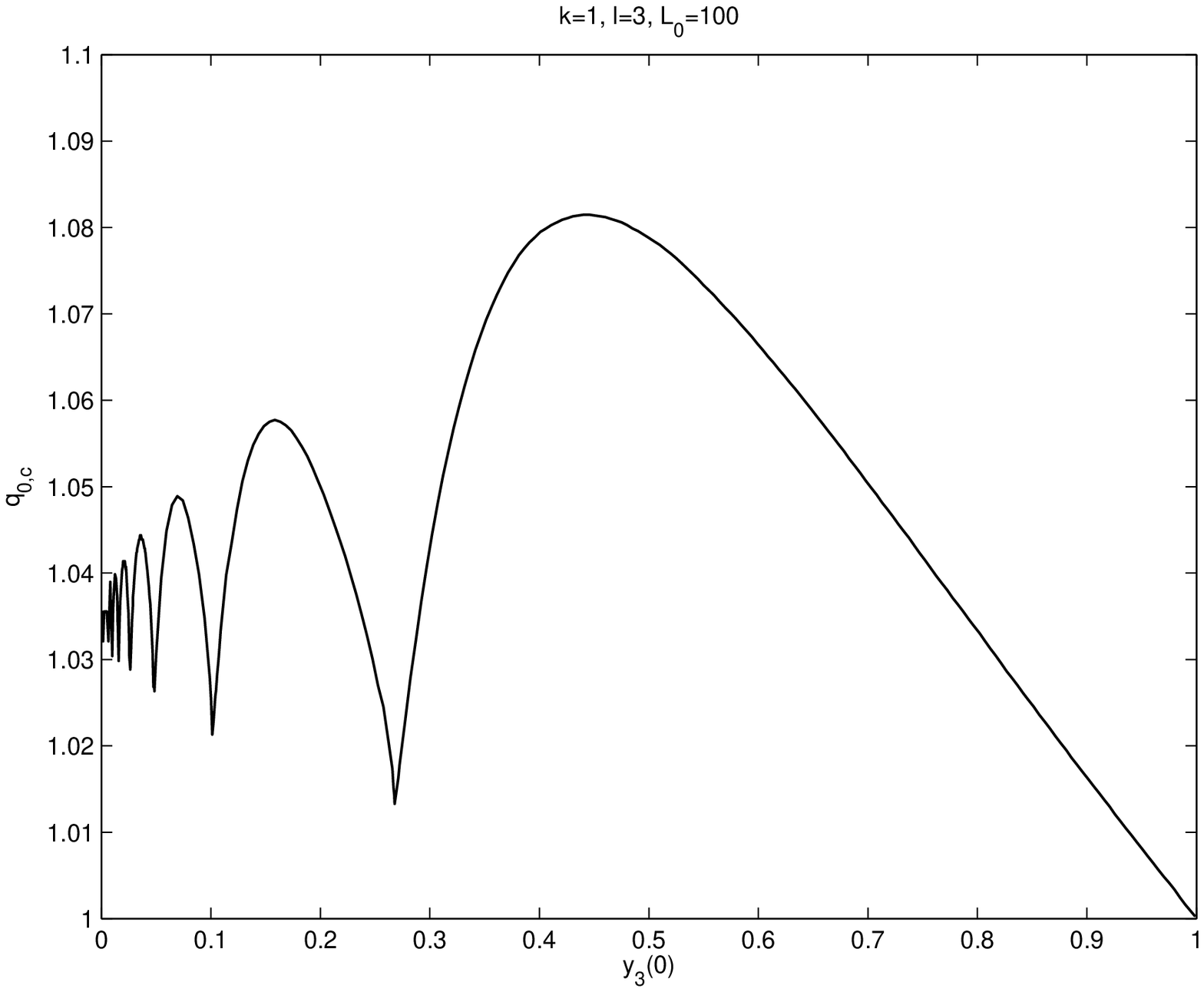}
\caption{$q_{0,c}$ as a function of $y_3(0)$}
\label{fig:q0c3}
\end{figure}

The value of $q_{0,c}$ for fixed values of $k, l$ and $L_0$ varies with
$y_3(0)$ as can be seen in Figure~\ref{fig:q0c3}.
We see that
$q_{0,c}(y_3(0))$
has an undulating quality with increasing amplitude and decreasing frequency
of oscillation for larger $y_3(0)$.
As $y_3(0) \to 1$, we do always have that $q_{0,c} \to 1$.
This is easily understood, since in the Newtonian case the critical
particle charge is exactly $1$, and in the limit $y_3(0) \to 1$
the solutions become essentially Newtonian; notice that
$1/y_3(0) -1$ is the central redshift factor \cite[Eqn.~(2.20)]{AR1}
which is a measure for how relativistic the system is.
At any point where $\rho$ equals zero, a Reissner-Nordström solution with
appropriate initial values can be joined to form a steady state with finite
support. The above limit for $q_0$ thus only applies to situations where the
distribution function is given by a single expression on the entire domain
interval $r \in [0,\infty[$.

Figure~\ref{fig:doublefig11} shows a double-peaked, single-shelled shell
configuration ($k=0, l=1.5, L_0=0.2, y_3(0)=0.05$). The positions and
magnitudes of peaks are virtually unaffected by the value of $q_0$ and the
only effect that can be seen is that the tail grows in length and
magnitude. The outer radius $R$ increases strictly monotonically. In
Figure~\ref{fig:doublefig08} a double-peaked ball configuration with parameters
($y_3(0)=0.01,k=1,l=12,L_0=0$) is displayed. Here we see that as $q_0$
increases, an extra maximum of $\rho$ appears. In this case $R$ does not
increase strictly monotonically, as a decrease can be seen before the outer
radius finally blows up as $q_0$ approaches $q_{0,c}$. Solutions in which a
new shell appears and with strictly increasing $R$ can however be
constructed. Although barely noticable in Figure~\ref{fig:doublefig08}, the
radial position of the new maximum that appears for larger $q_0$ will increase
at first and then decrease. This effect is more pronounced in
Figure~\ref{fig:doublefig12}, however.
The aforemensioned case is displaying the same behavior as in
Figure~\ref{fig:doublefig08},
this time for a single-peaked shell configuration
($y_3(0)=0.16,k=0,l=1.5,L_0=0.2$). For higher values of $L_0$ multi-shelled
configurations (i.e., configurations with multiple peaks, separated by vacuum
regions), can be obtained. The effect on these is that as $q_0$ increases, one
or more additional shells appear as can be seen in
Figure~\ref{fig:doublefig03} ($y_3(0)=0.0025,k=1,l=3,L_0=100$). These newly appearing
shells mimic the behavior of the newly appearing peak in
Figure~\ref{fig:doublefig12}. In Figure~\ref{fig:doublefig03} it can be seen clearly
that all peaks, except the innermost one, that are present at $q_0 = 0$ are
showing an inclination to move towards $r = 0$. This behavior is also present
in all cases with more than one peak, although not noticable in the plots.
The innermost peak, on the other hand, has a contrary tendency to move outwards.
\section{Sharpness issues of the main inequality}
\setcounter{equation}{0}

The purpose of this section is to investigate aspects concerning
sharpness of the inequality
\begin{equation}
\sqrt{m_g(r)} \leq \dfrac{\sqrt{r}}{3} + \sqrt{\dfrac{r}{9} +
\dfrac{q^2(r)}{3r}} \label{eq:aineq}
\end{equation}
where $m_g$ is the total gravitational mass given by
\begin{equation}\label{mg}
m_g(r) = 4\pi\int_0^r \eta^2\rho(\eta) d\eta +
\left(\int_0^r \dfrac{q^2(\eta)}{2\eta^2} d\eta + \dfrac{q^2(r)}{2r}\right)=:m_i+m_q.
\end{equation}
This inequality was derived in \cite{An2} and it was shown to hold
for any static solution of the Einstein-Maxwell-matter system
which satisfies (\ref{energy}). In addition it was assumed that
the solutions satisfy
\begin{equation}
q\leq m_g, \; m_g+\sqrt{m_g^2-q^2}<r.\label{inconditions}
\end{equation}
The latter conditions are imposed to ensure that the solutions are
physically meaningful, cf. \cite{GR}. To better understand the
motivation for our study we recall the results in the uncharged
case.

If $Q=0$ the inequality (\ref{eq:aineq}) reduces to the Buchdahl
inequality
\begin{equation}\label{B}
\frac{m(r)}{r}\leq\frac49,
\end{equation}
which was first proved in \cite{Bu1} under the Buchdahl assumptions
that the pressure is isotropic and the energy density is
non-increasing outwards. The inequality was then shown to hold
independently of the Buchdahl assumptions \cite{An3} for solutions
which satisfy the energy condition $p+2p_T\leq\rho.$ A different
proof was later given in \cite{KS}. The advantage of the latter
method is that the proof is shorter and more flexible since it
allows for other energy conditions than (\ref{energy}). The
disadvantage lies in the issues of sharpness and construction of the
saturating solution. Firstly, the method does not imply that the
class of saturating solutions is unique, and secondly, it is not
clear that a solution to a \textit{coupled} Einstein-matter system
can have the properties of the saturating solution constructed in
\cite{KS}. In particular solutions to the Einstein-Vlasov system
are ruled out. These issues have however affirmative answers.
Uniqueness is obtained in \cite{An3} where it is proved that a
saturating solution must be an infinitely thin shell solution. In
\cite{An4} it is shown that regular, arbitrarily thin shell solutions
of the Einstein-Vlasov system exist, which implies that there are
steady states to this system with $M/R$ arbitrarily close to $4/9.$

Let us now return to the charged case and discuss the latter two
issues. The proof of (\ref{eq:aineq}) in \cite{An2} is based on
the method in \cite{KS} and the method in \cite{An3} does not
apply. A proof based on the latter method would imply
uniqueness of the saturating solution, and it is thus natural to
ask if there is another class than infinitely thin shell solutions
with this property. Also, although the result in \cite{An2} shows
that infinitely thin shell solutions saturate the inequality there
is no analogous result to \cite{An4} in the charged case, i.e.,
the question whether or not the Einstein-Vlasov-Maxwell system
admits arbitrarily thin shell solutions is open. Below we will
present numerical evidence that the system does admit arbitrary
thin shell solutions and in addition that another class of
saturating solutions does exist.

\begin{figure}[Htpb]
\centering
    \includegraphics[scale=0.6]{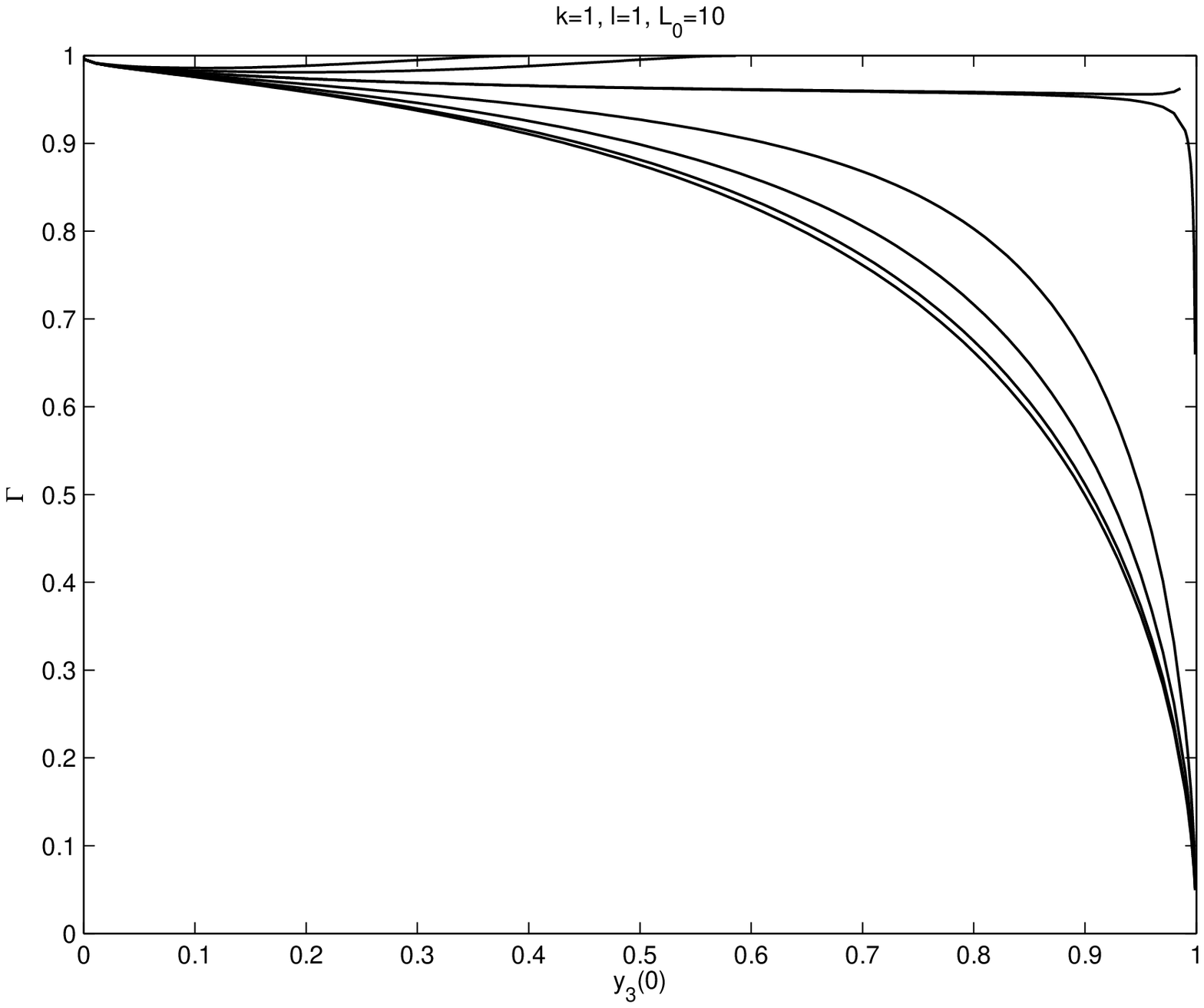}
\caption{$\Gamma$ as a function of $y_3(0)$ for a shell configuration}
\label{fig:aineq2}
\end{figure}

We introduce the quantity
\begin{equation*}
\Gamma := \sup_{r\geq0}
\dfrac{\sqrt{m_g(r)}}{\dfrac{\sqrt{r}}{3}+ \sqrt{\dfrac{r}{9} +
\dfrac{q^2(r)}{3r}}},
\end{equation*}
which in terms of the functions introduced in
\eqref{eq:ydef} reads
\begin{equation*}
\Gamma = \sup_{r\geq0}
\dfrac{3\sqrt{2}}{2}\dfrac{\sqrt{1-y_2^{-1}(r) + 4\pi
r^2y_1^2(r)}}{1 + \sqrt{1 + 12\pi r^2y_1^2(r)}}.
\end{equation*}
By \eqref{eq:aineq}, $\Gamma$ is subject to the inequality
\begin{equation}
\label{eq:gammaineq}
\Gamma \leq 1.
\end{equation}
In Figure~\ref{fig:aineq2} the numerical
results for a shell configuration with $(k=1,l=1,L_0=10)$ are displayed as follows.
$\Gamma$ as a function of
$y_3(0)$ is displayed in ascending order for $$q_0 \in \{0, 0.25, 0.5,
0.75, 0.999, 1.001, 1.25, 1.5\},$$ the bottom-most curve being that for $q_0
= 0$, the next to bottom-most being that for $q_0 = 0.25$ and so on.
In all cases tested for other parameter values, $\Gamma(y_3(0))$ falls within the first shell
of the solution.

\begin{figure}[Htpb]
\centering
    \includegraphics[scale=0.6]{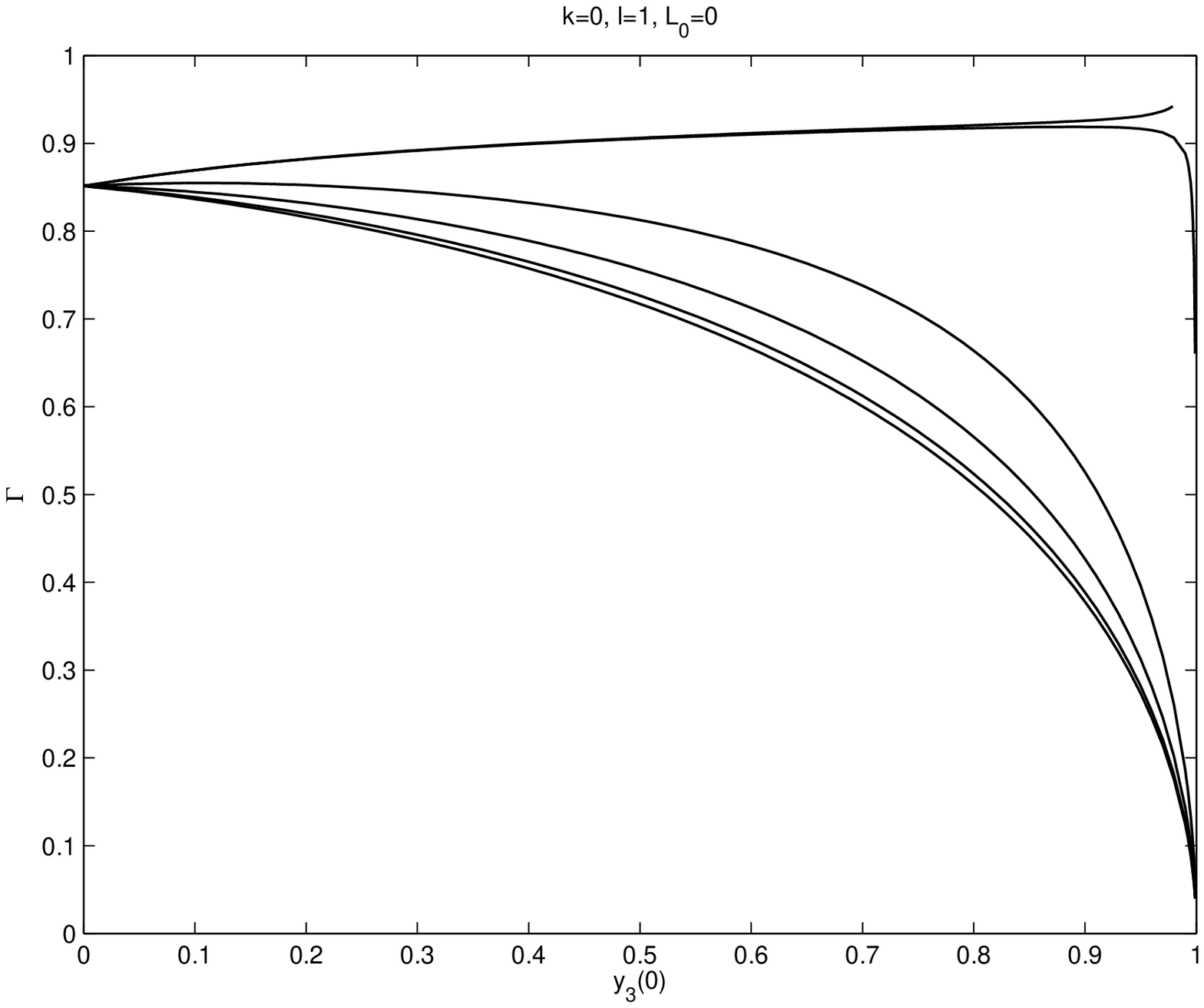}
\caption{$\Gamma$ as a function of $y_3(0)$ for a ball configuration}
\label{fig:aineq3}
\end{figure}

\begin{figure}[Htpb]
\centering
    \includegraphics[scale=0.6]{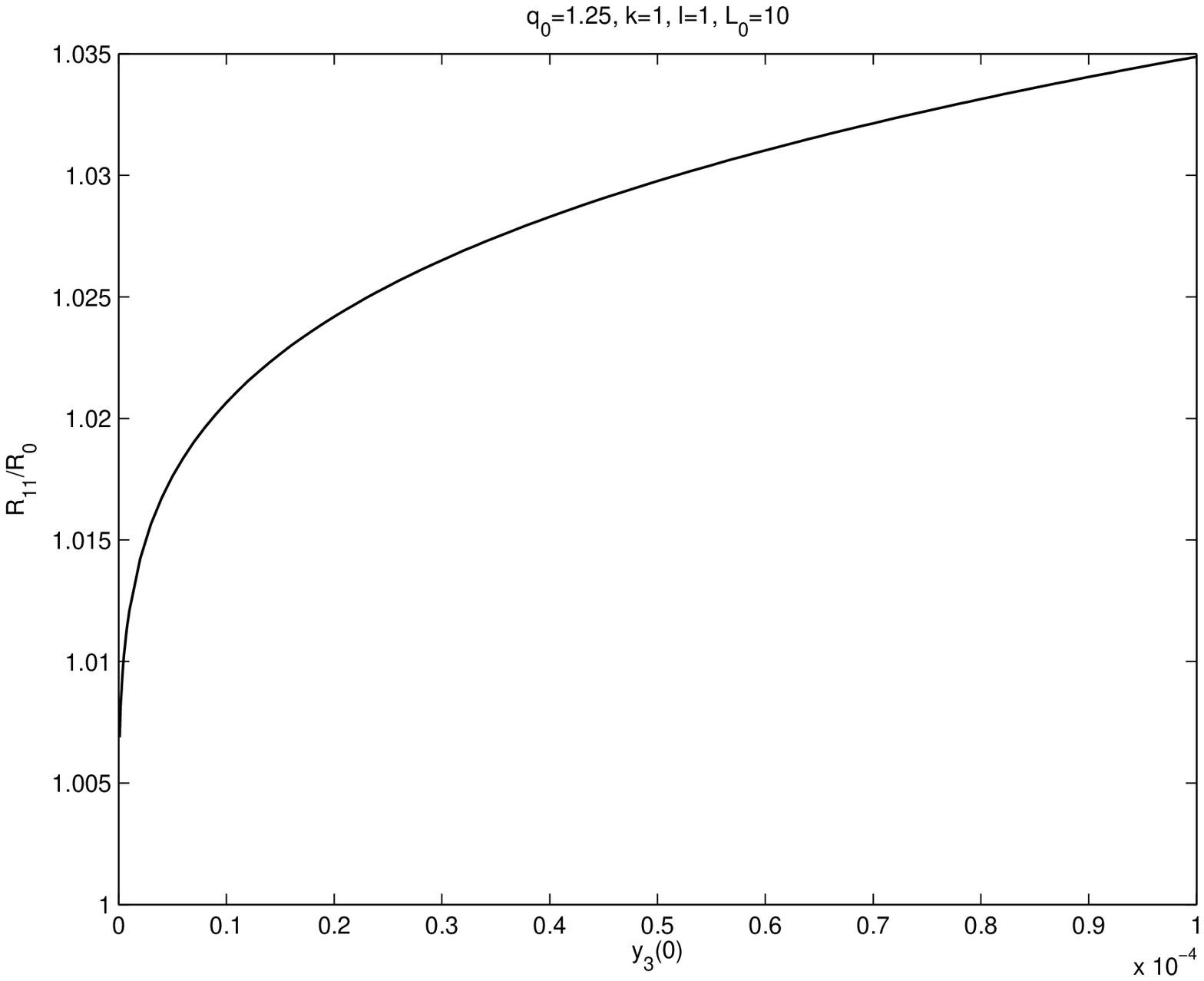}
    \caption{$R_{11}/R_0$ as a function of $y_3(0)$}
\label{fig:rquot}
\end{figure}

\begin{figure}[Htpb]
\centering
    \includegraphics[scale=0.6]{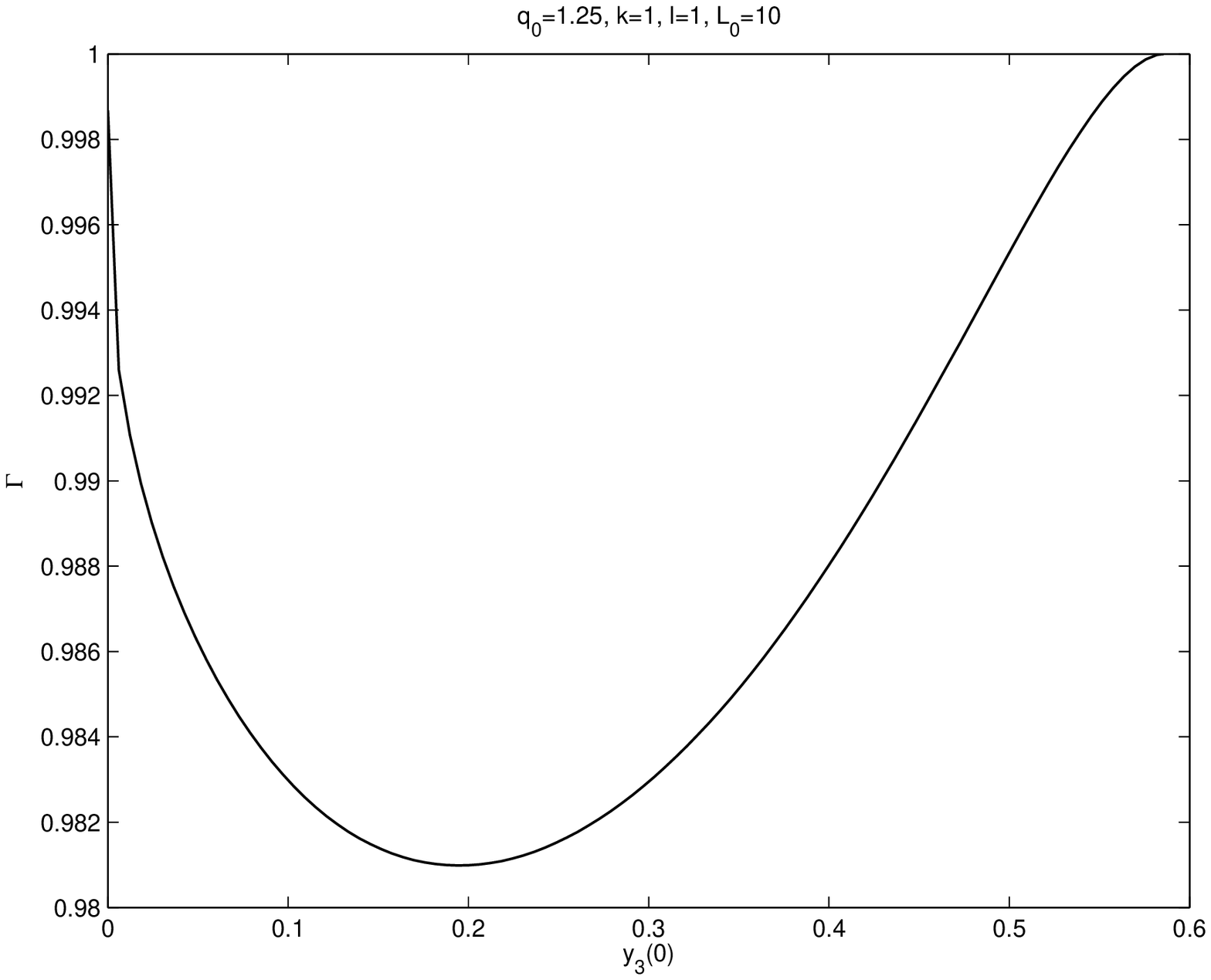}
\caption{$\Gamma$}
\label{fig:aineq4}
\end{figure}

\begin{figure}[Htpb]
\centering
    \includegraphics[scale=0.6]{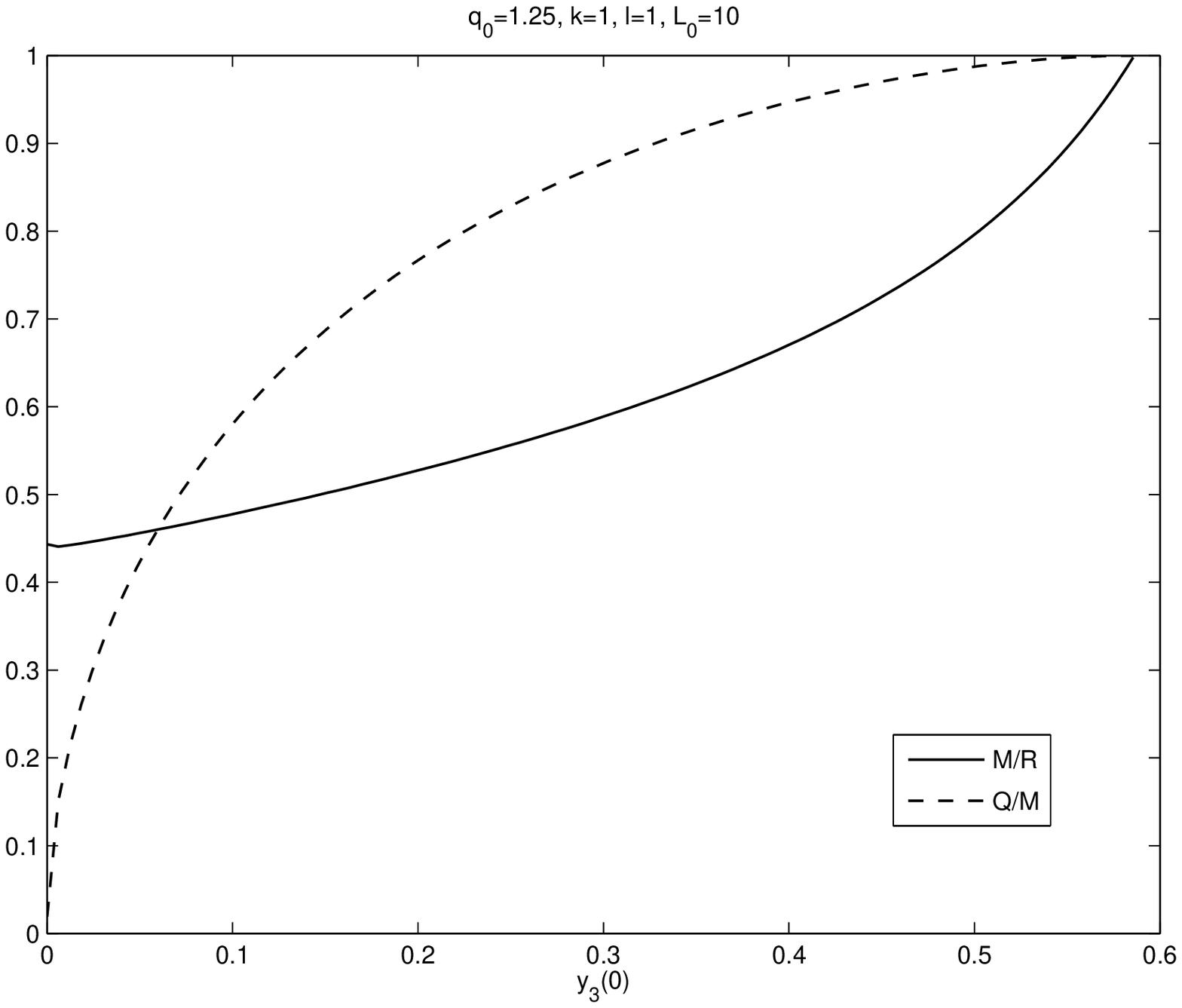}
\caption{$Q/M$ and $M/R$ as functions of $y_3(0)$}
\label{fig:quot}
\end{figure}

\begin{figure}[Htpb]
\centering
    \includegraphics[scale=0.6]{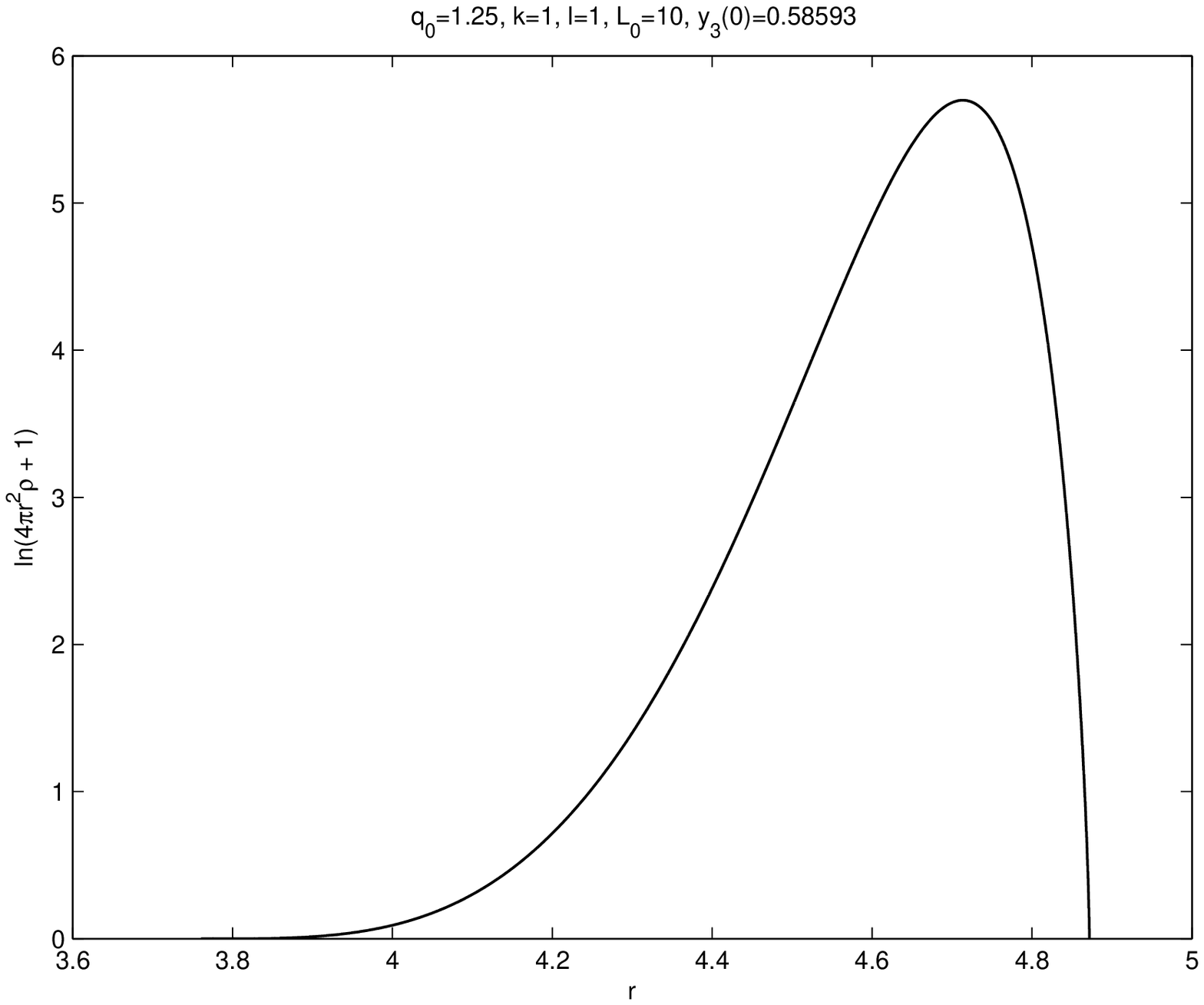}
\caption{The profile of a saturating solution which is not a thin shell}
\label{fig:notunique}
\end{figure}

\begin{figure}[Htpb]
\centering
    \includegraphics[scale=0.6]{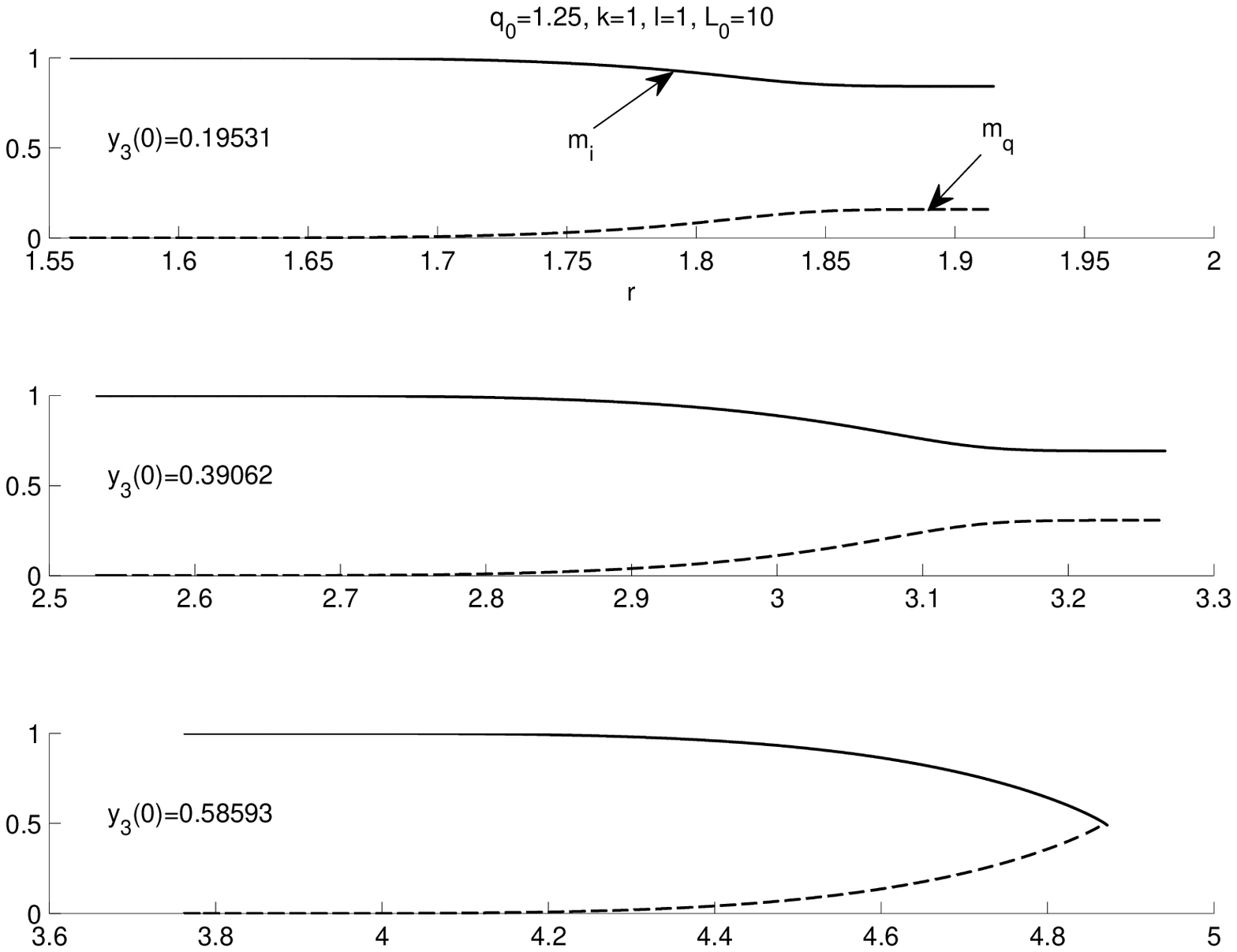}
\caption{The relation between $m_i$ and $m_q$}
\label{fig:mimq}
\end{figure}

We see that as $y_3(0)$ approaches zero \eqref{eq:gammaineq} approaches equality for all
values of $q_0$. By letting $R_{11}$ be the outer radius of the
first shell and plotting the ratio $R_{11}/R_0$ in Figure~\ref{fig:rquot},
we see that in the limit
$y_3(0) \to 0$, $R_{11}/R_0 \to 1$, i.e., we find numerical support that
the Einstein-Vlasov-Maxwell system admits arbitrarily thin shell solutions.
For values of $q_0 \leq 1$, we see that $\Gamma(y_3(0))$
is monotonically decreasing, and as $y_3(0)$ approaches unity,
$\Gamma$ approaches zero. This, however, is not the case for $q_0
> 1$ where $\Gamma(y_3(0))$ will at some point start
increasing. For values of $q_0$ slightly larger than $1,$
$\Gamma$ will not increase rapidly enough
for \eqref{eq:gammaineq} to once again be saturated.
For larger values of $q_0$, \eqref{eq:gammaineq} will
however be saturated a second time, cf.\ Figure~\ref{fig:aineq4} where the graph
for the case with $q_0=1.25$ is depicted using a different scale on the $\Gamma-$axis.
From Figure~\ref{fig:quot} it is clear that this
occurs when $Q=M=R$. Hence, in the charged case the class of
saturating solutions is not unique. Figure~\ref{fig:notunique}
displays the graph of $\rho$ for a solution which nearly saturates the
inequality and such that $M,Q$ and
$R$ are almost equal, and we see that it is indeed not a thin shell solution.
In equation (\ref{mg}) the quantities $m_i$ and $m_q$ were defined, and roughly
they represent the parts of the gravitational mass induced by $\rho$ and $q$ respectively.
However, it should be noted that the nonlinearity of the equations make it impossible to
completely separate the influences of these quantities. In figure~\ref{fig:mimq} these
quantities are plotted in three different cases. Note in particular that in the third case,
which corresponds to the saturating solution for which $M=Q=R,$ $m_i$ and $m_q$ are almost
equal at the outer boundary of the solution.

Figure~\ref{fig:aineq3} shows $\Gamma(y_3(0))$ for the family of
ball configurations with $(k=0,l=1,L_0=0)$, for $q_0=\{0,0.25,0.50,0.75,0.999,1.001\}$.

Although no longer
monotonically decreasing, $\Gamma(y_3(0))$ approaches zero as
$y_3(0)$ approaches unity for $q_0 < 1$, as in the case for
the above shell configurations.

\section{Spirals in the mass radius diagram}
\begin{figure}[Htpb]
\centering
    \includegraphics[scale=0.6]{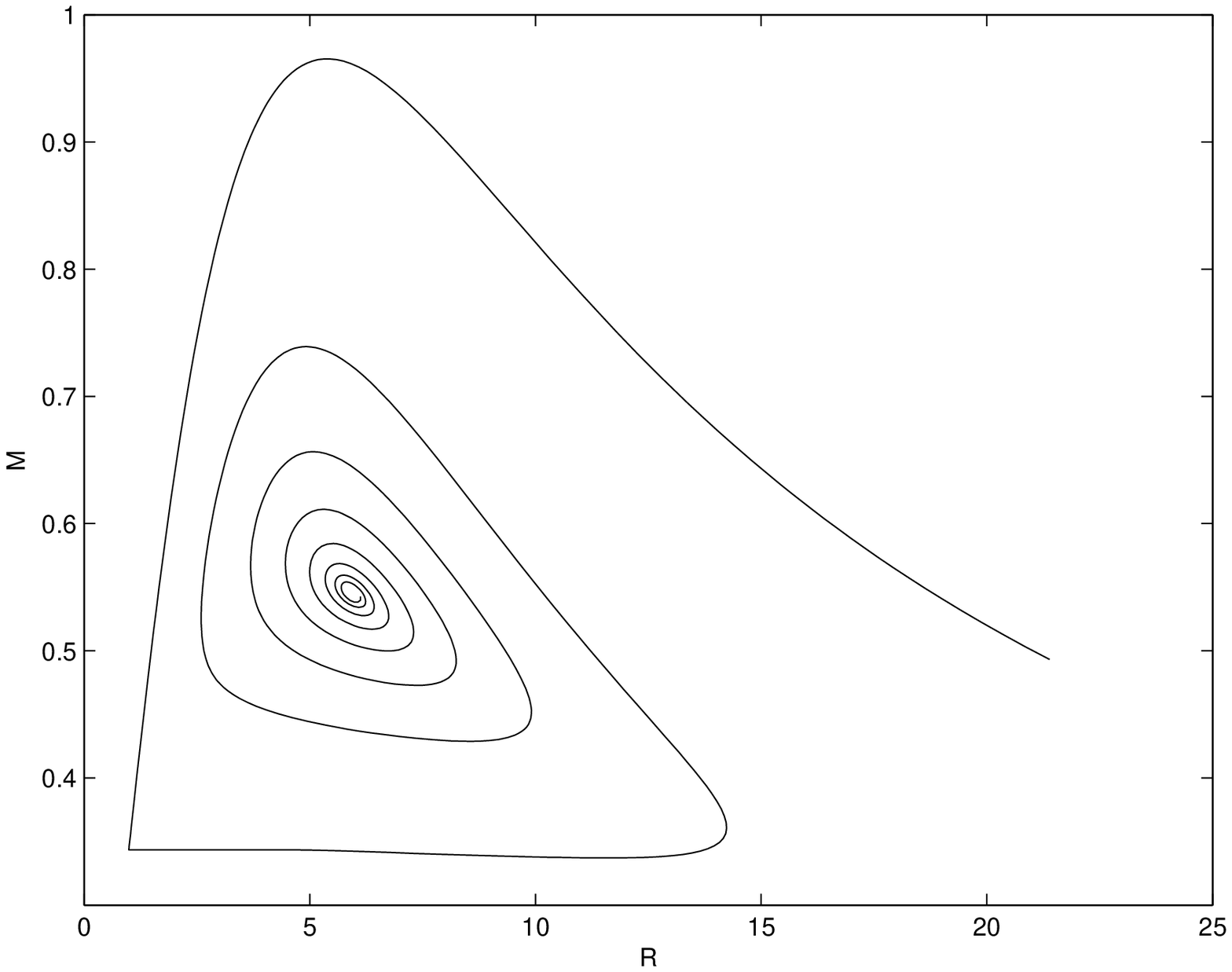}
   \caption{$(R,M)$-spiral, $q_0=0$}
\label{fig:spiral0}
\end{figure}

\begin{figure}[Htpb]
\centering
  \includegraphics[scale=0.6]{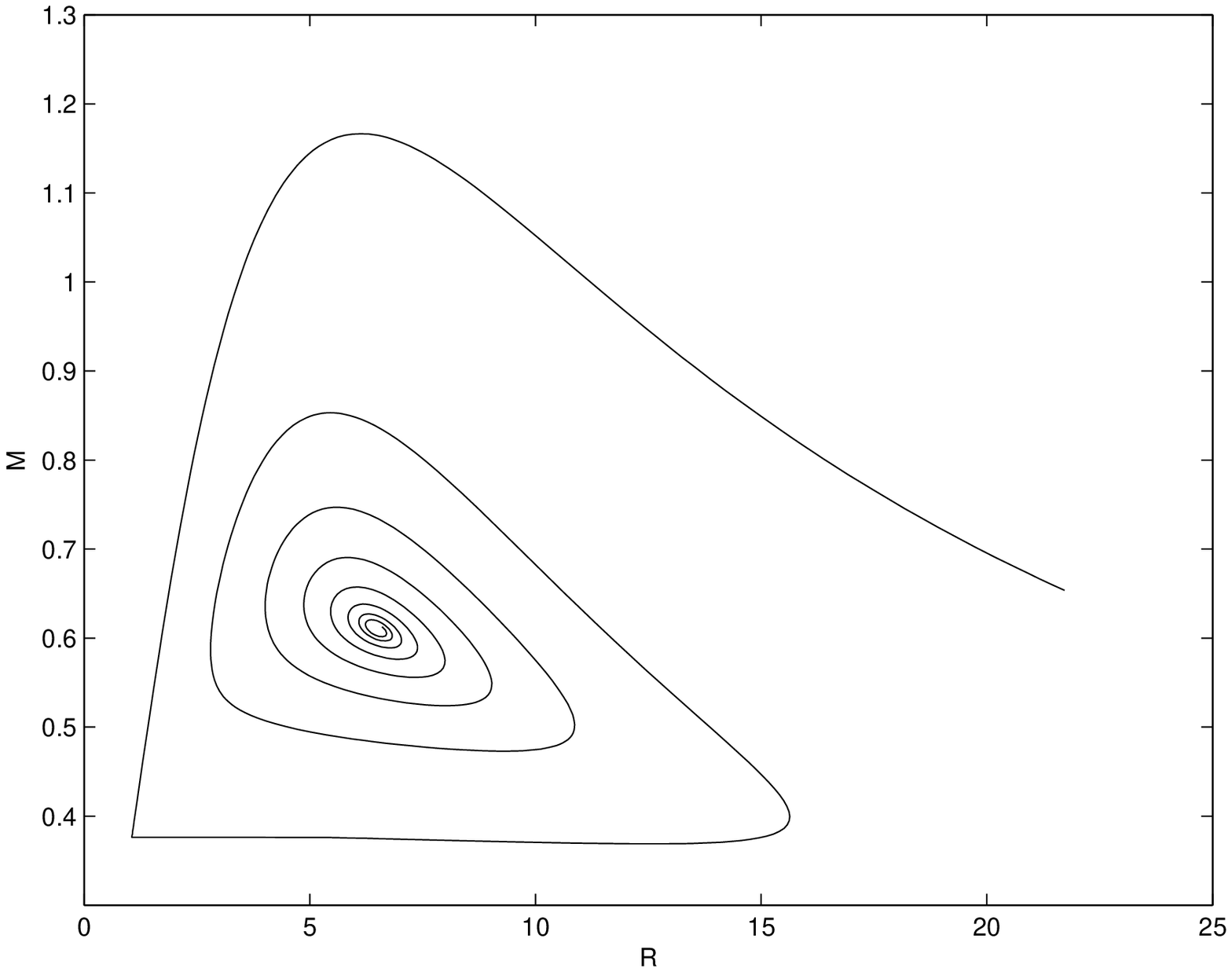}
  \caption{$(R,M)$-spiral, $q_0=0.5$}
\label{fig:spiralh}
\end{figure}

\setcounter{equation}{0}
In this section we study the behavior of the total gravitational
mass $M$ and outer radius of the support $R$ for one parameter
families of steady states. The parameter is $y_3(0)$ while $k, l, L_0$ and
$q_0$ are kept constant. In \cite{AR1} it has been shown that
in the isotropic (i.e., $l=L_0=0$) and chargeless case, $(R,M)$
forms a spiral. Furthermore, numerical evidence is given that this
is also the case for non-isotropic solutions. Charged matter, as
studied in this paper, displays the same behavior and changing
$q_0$ merely deforms the spirals. This can be seen in
Figures~\ref{fig:spiral0} and \ref{fig:spiralh} showing the
$(R,M)$-spirals for $q_0=0$ and $q_0=0.5$ with ($k=1,l=5,L_0=2$).
We see that increasing
$q_0$ from $0$ to $0.5$ does not change the shape
significantly. These characteristics are displayed for all combinations of
$k,l$ and $L_0$.
The sharp corner in these mass-radius spirals is a genuine feature,
since for $L_0>0$ the radius of the support as $y_3(0)$ varies can change
discontinuosly  due to new shells which appear,
cf.\ \cite[p.~1829]{AR1}.

\begin{figure}[Htpb]
\centering
    \includegraphics[scale=0.6]{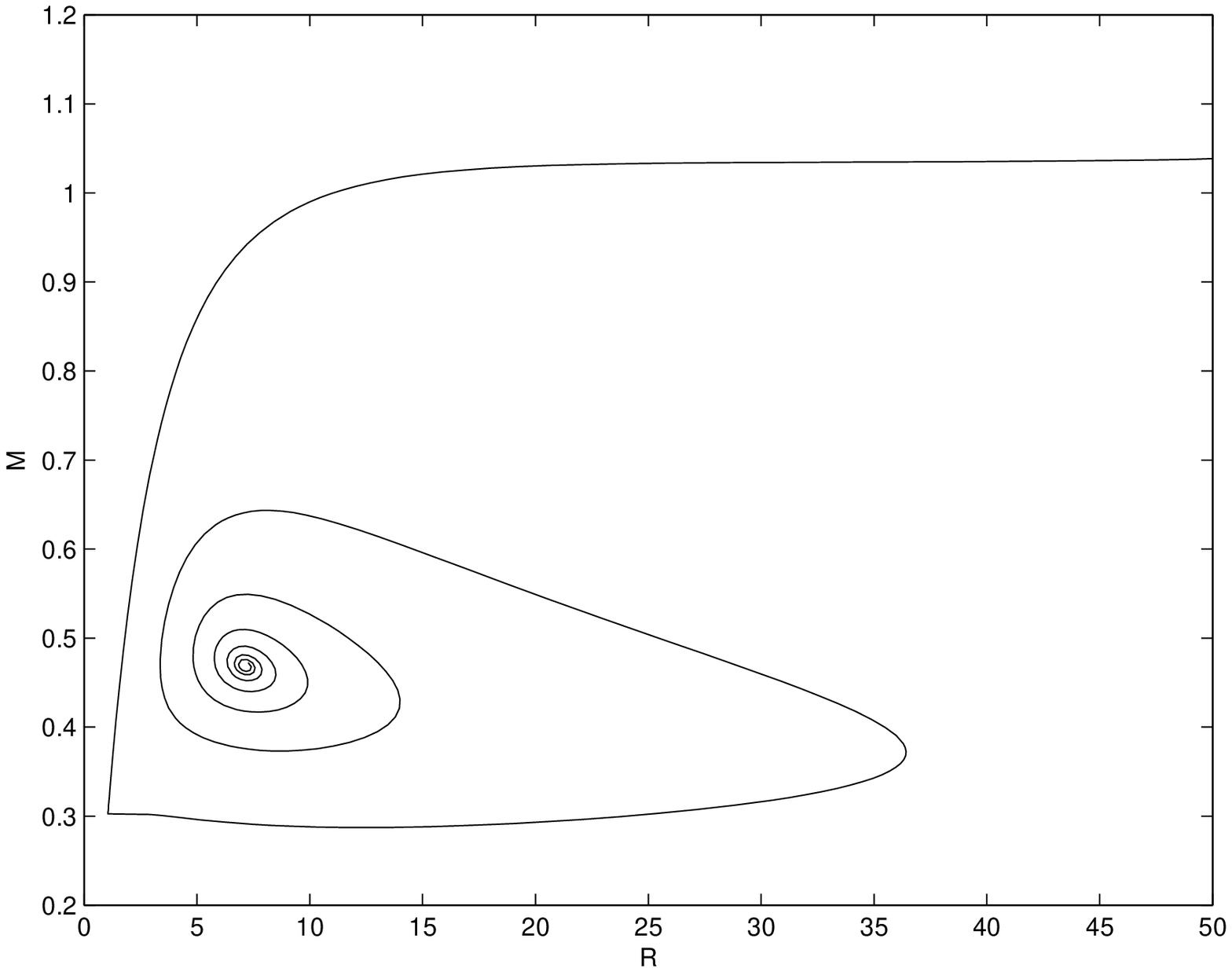}
  \caption{($R,M)$-spiral, $q_0=0.5$, alternative ansatz}
\label{fig:sin2spiral}
\end{figure}

So far in this paper we have only used the ansatz
\eqref{eq:numansatz}for the distribution function. Only the
analogous ansatz is used in \cite{AR1}, so it might be
interesting to see if the $(R,M)$-spirals are also a feature of
other ansatz functions of the form $f(r,w,L) = \phi(E/E_0,L)$. For
instance, with $\phi(E/E_0,L) =
\sin^2\left(k(1-E/E_0)_+\right)\sin^2\left(l(L-L_0)_+\right)$,
and ($k=2, l=3, L_0=1$, $q_0=0.5$), we get the spiral in
Figure~\ref{fig:sin2spiral}. For all cases tested, we do in fact
get $(R,M)$-spirals. We conclude that $(R,M)$-spirals are a
general feature of the Einstein-Vlasov-Maxwell system with the
ansatz $f(r,w,L) = \phi(E/E_0,L)$.

\end{document}